\documentclass[10pt,journal,twocolumn]{IEEEtran}
\IEEEoverridecommandlockouts
\usepackage{algpseudocode}
\usepackage{algorithm}
\usepackage{bm}
\usepackage{amsfonts}
\usepackage{amssymb}
\usepackage{amscd}
\usepackage{graphics}
\usepackage[dvips]{graphicx}
\usepackage{epsfig}
\usepackage{subfigure}
\usepackage[cmex10]{amsmath}
\usepackage{array}
\usepackage{eqparbox}
\usepackage{flushend}
\usepackage{hyperref}
\usepackage{fancyhdr}
\usepackage{titling}
\usepackage{enumerate}

\DeclareMathAlphabet{\mathbsf}{OT1}{cmss}{bx}{n}
\DeclareMathAlphabet{\mathssf}{OT1}{cmss}{m}{sl}
\DeclareMathAlphabet{\mathcsf}{OT1}{cmss}{sbc}{n}

\newcommand{\ie}{{\em i.e.}}
\newcommand{\etc}{{\em etc}}

\newcommand{\secref}[1]{Section~\ref{#1}}
\newcommand{\figref}[1]{Fig.~\ref{#1}}
\newcommand{\tabref}[1]{Table~\ref{#1}}

\newcommand{\appref}[1]{Appendix~\ref{#1}}

\makeatletter
\def\blfootnote{\xdef\@thefnmark{}\@footnotetext}
\makeatother

\renewcommand{\baselinestretch}{1}
\newcommand{\qed}{\nobreak \ifvmode \relax \else
      \ifdim\lastskip<1.5em \hskip-\lastskip
      \hskip1.5em plus0em minus0.5em \fi \nobreak
      \vrule height0.75em width0.5em depth0.25em\fi}

\def\BibTeX{{\rm B\kern-.05em{\sc i\kern-.025em b}\kern-.08em
    T\kern-.1667em\lower.7ex\hbox{E}\kern-.125emX}}

\usepackage{newlfont}

\begin{document}
\title{Electrical Structure-Based PMU Placement in Electric Power Systems}
\author{\IEEEauthorblockN{K. G. Nagananda}\thanks{The author was with the Department of Electrical and Computer Engineering, Lehigh University, Bethlehem, PA $18015$, U.S.A. E-mail: \texttt{kgn209@lehigh.edu}}
}

\pagenumbering{gobble}
\date{}
\maketitle

\begin{abstract}
Recent work on complex networks compared the topological and electrical structures of the power grid, taking into account the underlying physical laws that govern the electrical connectivity between various components in the network. A distance metric, namely, resistance distance was introduced to provide a more comprehensive description of interconnections in power systems compared with the topological structure, which is based only on geographic connections between network components. Motivated by these studies, in this paper we revisit the phasor measurement unit (PMU) placement problem by deriving the connectivity matrix of the network using resistance distances between buses in the grid, and use it in the integer program formulations for several standard IEEE bus systems. The main result of this paper is rather discouraging: more number of PMUs are required, compared with those obtained using the topological structure, to meet the desired objective of complete network observability without zero injection measurements. However, in light of recent advances in the electrical structure of the grid, our study provides a more realistic perspective of PMU placement in power systems. By further exploring the connectivity matrix derived using the electrical structure, we devise a procedure to solve the placement problem without resorting to linear programming.
\end{abstract}


\section{Introduction}\label{sec:introduction}
The phasor measurement unit (PMU) is a critical component of today's energy management systems, designed to enable near-realtime wide area monitoring and control of the electric power system. Synchrophasor data are used for applications varying from state estimation, islanding control, identifying outages, voltage stability detection and correction, disturbance recording, {\etc}. An operational feature of the PMU is to record not only the voltage magnitude and phase angles at the bus where it is installed, but also the current phasor of all lines incident on that bus. This results in a highly correlated set of measurements at each PMU, leading to the following question: given the redundancy in phasor measurements, what is the minimum (or optimal) number of PMUs to be placed in the power network so as to make the system measurement model observable, and thereby linear? This is commonly referred to as the PMU placement problem, and is well reported in the literature (see \cite{Baldwin1993} - \nocite{Milosevic2003} \nocite{Xu2004}\nocite{Nuqui2005}\nocite{Chen2006} \nocite{Gou2008}\nocite{Gou2008a}\nocite{Zhang2010}\nocite{Deka2011}\nocite{Azizi2012}\nocite{Kekatos2012}\cite{Li2013}).

\subsection{Literature review}\label{subsec:literature_review}
Typically, the PMU placement problem is cast in the mathematical programming framework, and various strategies have been investigated to obtain the optimal number of PMUs to satisfy the desired optimization criterion. The most commonly employed criteria are complete and incomplete network observability \cite{Nuqui2005}, with and without zero power injection measurements. For instance, in \cite{Baldwin1993}, the placement was provided by a spanning measurement subgraph, and the minimal PMU set was obtained using a dual search algorithm incorporating modified bisecting simulated-annealing-based searches for complete network observability. In \cite{Milosevic2003}, a Pareto-optimal solution was obtained for PMU placement using non-dominated sorting genetic algorithm. However, in the presence of power injection measurements, the proposed integer program turned out to be nonlinear. In \cite{Xu2004}, the problem was formulated as an integer program to include conventional power flow and injection measurements in addition to PMU measurements for maximum network observability. In \cite{Chen2006}, the objective was to achieve bad data detection during state estimation using optimal PMU placement. In \cite{Gou2008},\cite{Gou2008a}, optimal PMU locations were obtained by formulating the problem as an integer linear program for complete and incomplete network observability, with and without conventional power flow and injection measurements.

In \cite{Zhang2010}, optimal PMU placement was linked with power system dynamic state estimation. A method for evaluating a specific PMU placement, when there are multiple placement solutions, was also proposed. A computationally efficient distributed algorithm using affinity propagation framework to solve the integer linear program was proposed in \cite{Deka2011}; within this framework, communications between nodes in the grid was utilized to also perform error correction for PMUs that suffered from measurement errors. In \cite{Azizi2012}, the integer linear program was solved using an exhaustive search-based method with complete network observability, and the state estimation implemented on such a placement was shown to be linear. A convex relaxation method was employed in \cite{Kekatos2012} to optimize the PMU placement based on estimation-theoretic criteria, where state estimation is accomplished under the Bayesian framework. However, studies in the above references are based on the topological structure or node connectivity (degree) of the power network.

\subsection{Electric power grid: A complex networks perspective}\label{subsec:complex_networks}
Given its size and societal importance, the electric power grid has received considerable attention from the perspective of complex networks \cite{Dorfler2010}. Studies showed that, for many classes of complex networks, characterizing the network structure using degree distribution alone was suboptimal, and had implications on node synchronization and performance of the network (for instance, see \cite{Wu1995} - \nocite{Wu2005}\cite{Atay2006}). In the context of the electric power grid, it was reported that grids in different geographical regions had different degree distributions, leading to varied topological structures. Furthermore, it was also shown that different model based analyses of the same power grid had resulted in different topological structures. The reason for this discrepancy is that topological structures are solely based on geographic separation and neglect the underlying physical laws (Ohm's and Kirchoff's) that govern the electrical connections or flows between network components. These issues were illuminated in \cite{Cotilla-Sanchez2012} (see Section I and references therein), where the topological and electrical structures of the electric grid were compared and analyzed. The work in \cite{Cotilla-Sanchez2012} showed how the power grid differed from the commonly employed complex networks used to model the structure of the grid, namely, random graphs \cite{Erdos1959}, small-world networks \cite{Watts1998} and preferential attachment graphs \cite{Barabasi1999}. To study the electrical structure, \cite{Cotilla-Sanchez2012} (see also \cite{Wang2010b}) introduced the electrical centrality measure to characterize the connectivity or betweenness of nodes in the power network. A distance metric, namely, resistance distance was introduced to provide a more comprehensive description of interconnections between components compared with the topological structure, which is based only on degree distributions or direct physical connections.

\subsection{Electrical structure-based PMU placement}\label{subsec:elecstruc_pmu}
On the one hand, we have a significant problem of PMU placement whose solution is based on the structure of the power grid, while on the other hand, there are results in the area of complex networks which promote the electrical structure of the power network, over its topological structure. Therefore, a natural question that arises is the following: what implications do the electrical structure have on PMU placement in the power grid? We investigate this question in this paper.

The first step in this direction is to derive the connectivity or adjacency matrix - used in the the PMU placement problem formulation - using resistance distances rather than direct physical connections between buses in the power grid. Next, we formulate integer programs in this manner for several standard IEEE test bus systems, without conventional measurements and for complete network observability. The main result of our work is rather discouraging: more number of PMUs, compared with those obtained using the topological structure, are required to meet the objective of complete network observability. However, in light of the recent advances in the electrical structure of the grid, the results presented in this paper provide a more realistic perspective of PMU placement in power systems. We introduce the notion of average resistance distance and use it in conjunction with the graph of the connectivity matrix, obtained using the electrical structure, to devise a strategy for solving the PMU placement problem without resorting to integer linear programming. To the best of the author's knowledge, this is the first instance of the PMU placement problem being addressed from the perspective of the electrical structure of the power network.

The remainder of the paper is organized as follows. In \secref{sec:review_elecstruc}, we review some details of the electrical structure of the power grid, and build the necessary framework to be used in the rest of the paper. In \secref{sec:placement_problem}, we formulate the PMU placement problem based on the electrical structure, and tabulate the main results of the paper. Further insights into the placement problem are provided in \secref{sec:discussion}. We conclude the paper in \secref{sec:conclusion}.

\section{Electrical Structure of the Power Network: A Review }\label{sec:review_elecstruc}
In this section, we review the electrical structure of the power network. The concept of resistance distance is introduced, and the procedure to derive the binary connectivity matrix of a given power grid is presented. The exposition follows directly from \cite[Section III]{Cotilla-Sanchez2012}, and is presented here for sake of completeness and clarity.

The sensitivity between power injections and nodal phase angles differences can be utilized to characterize the electrical influence between network components. The electrical structure of the power network can then be understood by measuring the amount of electrical influence that one component has on another in the network. The measurement of this electrical influence necessitates a metric system. Mathematically, this can be accomplished by first deriving the sensitivity matrix, which can be obtained by standard methods. The complement of the sensitivity matrix is called the distance matrix, whose entries quantify the electrical influence that each component has on the other - zero value indicates that two components are perfectly connected, while a large number indicates that the corresponding components have negligible electrical influence on each other. This electrical distance was proved to be a formal distance metric, and was employed to address various problems in power systems.

Another method to measure the electrical influence between network components is to derive the resistance distance \cite{Klein1993}, which is the effective resistance between points in a network of resistors. Consider a network with $N$ nodes, described by the conductance matrix $\bm{G}$. Let $V_j$ and $g_{ij}$ denote the voltage magnitude at node $j$ and the conductance between nodes $i$ and $j$, respectively. The current injection at node $i$ is then given by
\begin{eqnarray}\label{eq:current_injection}
I_i = \sum_{j=1}^{N}g_{ij}V_j.
\end{eqnarray}
$\bm{G}$ acts as a Laplacian matrix to the network, provided there are no connections to the ground, {\ie}, if $\bm{G}$ has rank $N-1$. The singularity of $\bm{G}$ can be overcome by letting a node $r$ have $V_r = 0$. The conductance matrix associated with the remaining $N-1$ nodes is full-rank, and thus we have
\begin{eqnarray}\label{eq:nonreferencenodes}
\bm{V}_k = \bm{G}^{-1}_{kk}\bm{I}_k, k \neq r.
\end{eqnarray}
Let the diagonal elements of $\bm{G}^{-1}_{kk}$ be denoted $g^{-1}_{kk}$, $\forall k$, indicating the change in voltage due to current injection at node $k$ which is grounded at node $r$. The voltage difference between a pair of nodes $(i,j)$, $i\neq j\neq r$, is computed as follows:
\begin{eqnarray}\label{eq:voltage_difference}
e(i,j) = g^{-1}_{ii} + g^{-1}_{jj} - g^{-1}_{ij} - g^{-1}_{ji},
\end{eqnarray}
indicating the change in voltage due to injection of $1$ Ampere of current at node $i$ which is withdrawn at node $j$. $e(i,j)$ is called the resistance distance between nodes $i$ and $j$, and describes the sensitivity between current injections and voltage differences. In matrix form, letting $\boldsymbol{\Gamma} \triangleq \text{diag}(\bm{G}^{-1}_{kk})$, we have $\forall k \neq r$
\begin{eqnarray}\label{eq:matrix_form}
\bm{E}_{kk} &=& \boldsymbol{1}\boldsymbol{\Gamma}^{\mathrm{T}} + \boldsymbol{\Gamma}\boldsymbol{1}^{\mathrm{T}} - \bm{G}^{-1}_{kk} - \left[\bm{G}^{-1}_{kk}\right]^{\mathrm{T}},\\
\bm{E}_{rk} &=& \boldsymbol{\Gamma}^{\mathrm{T}},\\
\bm{E}_{kr} &=& \boldsymbol{\Gamma}.
\end{eqnarray}
The resistance distance matrix $\bm{E}$, thus defined, possesses the properties of a metric space \cite{Klein1993}.

To derive the sensitivities between power injections and phase angles, we start with the upper triangular part of the Jacobian matrix obtained from the power flow analysis, for the distance matrix to be real-valued:
\begin{eqnarray}\label{eq:upper_jacobian}
\Delta \bm{P} = \left[\frac{\partial P}{\partial \theta}\right]\Delta \theta + \left[\frac{\partial P}{\partial |V|}\right]\Delta |V|.
\end{eqnarray}
The matrix $\left[\frac{\partial P}{\partial \theta}\right]$ will be used to form the distance matrix, by assuming the voltages at the nodes to be held constant, {\ie}, $|V|=0$. It was observed that $\left[\frac{\partial P}{\partial \theta}\right]$ possesses most of the properties of a Laplacian matrix. By letting $\bm{G} = \left[\frac{\partial P}{\partial \theta}\right]$, the resulting distance matrix $\bm{E}$ measures the incremental change in phase angle difference between two nodes $i$ and $j$, $(\theta_i - \theta_j)$, given an incremental average power transaction between those nodes, assuming the voltage magnitudes are held constant. It was proved in \cite[Appendix]{Cotilla-Sanchez2012} that $\bm{E}$, thus defined, satisfies the properties of a distance matrix, as long as all series branch reactance are nonnegative.

For a power grid with $N$ buses, the distance matrix $\bm{E}$ translates into an undirected graph with $N(N-1)$ weighted branches. In order to compare the grid with an undirected network without weights, one has to retain the $N$ buses, but replace the $M$ branches with $M$ smallest entries in the upper or lower triangular part of $\bm{E}$. This results in a graph of size $\{N,M\}$ with edges representing electrical connectivity rather than direct physical connections. The adjacency matrix $\bm{B}$ of this graph is obtained by setting a threshold, $\tau$, adjusted to produce exactly $M$ branches in the network:
\begin{eqnarray}\label{eq:adjcency_matrix}
\bm{B}:
\begin{cases}
b_{ij} = 1, ~\forall e(i,j) < \tau,\\
b_{ij} = 0, ~\forall e(i,j) \geq \tau
\end{cases}
\end{eqnarray}
In this paper, we will derive the binary connectivity matrix $\bm{B}$ for several standard IEEE test bus systems, for use in the PMU placement problem.

\section{PMU placement and main results}\label{sec:placement_problem}
In this section, we revisit the PMU placement problem with two different perspectives, one based on the topological structure of the power network and the other based on the electrical structure. In both cases, the problem is formulated as an integer linear program with one inequality constraint.

Given a power network with $N$ buses and $M$ branches, we only consider complete network observability without conventional measurements. We let $\bm{A}$ denote the binary connectivity matrix of dimensions $N\times N$, $\bm{x}$ with dimensions $N\times 1$ denote the binary decision variable vector defined as follows:
\begin{eqnarray}\label{eq:binarydecision_vector}
x_i = \begin{cases}
1, ~\text{if a PMU is installed at bus}~i,\\
0, ~\text{otherwise},
\end{cases}
\end{eqnarray}
where $i=1,\dots,N$, and $\bm{b}$ is a unit vector of dimensions $N\times 1$. The PMU placement problem is formulated as follows:
\begin{eqnarray}\label{eq:pmuplacement}
\nonumber \min \sum_{i=1}^{N}x_i\\
\text{such that}~ \bm{A}\bm{x} \geq \bm{b}\\
\nonumber x_i \in \{0,1\}
\end{eqnarray}

\subsection{PMU placement based on topological structure}\label{subsec:pmuplacement_topology}
For the case based on the topological structure of the power network, we consider the existing approach of deriving the binary connectivity matrix directly from the bus admittance matrix. The entries of the bus admittance matrix are transformed into binary form, and used in the problem setup \eqref{eq:pmuplacement}. The entries of $\bm{A}$ are given by
\begin{eqnarray}\label{eq:topological_A}
\bm{A}:
\begin{cases}
a_{ij} = 1, ~\text{if}~i=j,\\
a_{ij} = 1, ~\text{if}~i~\text{and}~j~\text{are connected},\\
a_{ij} = 0, ~\text{if}~i~\text{and}~j~\text{are not connected}.
\end{cases}
\end{eqnarray}
The entries $a_{ij}$ of $\bm{A}$ characterize the direct physical or geographic separation between the network components, without taking into account the electrical properties of these connections.

\subsection{PMU placement based on electrical structure}\label{subsec:pmuplacement_electrical}
For the electrical structure-based PMU placement, the binary connectivity matrix is derived as discussed in \secref{sec:review_elecstruc}, {\ie}, $\bm{A} = \bm{B}$ (as defined in \eqref{eq:adjcency_matrix}), and this will be used in the formulation \eqref{eq:pmuplacement}. As discussed in \secref{sec:review_elecstruc}, the entries $b_{ij}$ of $\bm{B}$ are obtained taking into account the physical laws that govern the electrical properties of the network connections.

\subsection{Experiments and results}\label{subsec:main_results}
We now present the experimental setup and the main results of this paper. The goal was to obtain the minimum number of PMUs by solving the placement problem based on both the topological and electrical structures of the power network, for several standard IEEE test bus systems. The criterion for the optimization problem was to achieve complete network observability, without conventional measurements. The bus and branch data, required to derive the bus admittance and power-flow Jacobian matrices, were obtained using MATPOWER \cite{Zimmerman2011} and archived resources \cite{Washington}. The binary integer programming tool of Matlab was used to solve the problem defined by \eqref{eq:pmuplacement}. In \tabref{tab:bus_pmu}, we tabulate the main results of this paper.
\begin{table}[h]
\centering
\begin{tabular}{ | c | c | c |}\hline
    IEEE bus system & Topological structure & Electrical structure \\ \hline
    9 & 3 &  4\\ \hline
    14 & 4 & 7 \\ \hline
    30 & 10 & 17 \\ \hline
    39 & 13 & 22 \\ \hline
    57 & 17 & 35 \\ \hline
    118 & 32 & 93 \\ \hline
    162 & 43 & 125 \\ \hline
    \end{tabular}
    \caption{Minimum number of PMUs based on topological and electrical structures for IEEE test bus systems}
    \label{tab:bus_pmu}
\end{table}
As shown in \tabref{tab:bus_pmu}, for each test bus system, the minimum number of PMUs obtained using the electrical structure is more than that obtained using the topological structure of the network, to meet the objective of complete network observability without zero injection measurements.

\section{Discussion}\label{sec:discussion}
In this section, we provide further insights into the placement problem. Interestingly, we identify an avenue to obtain the minimum number of PMUs \emph{without} formulating an integer program (as in \eqref{eq:pmuplacement}). This also leads to a better understanding of the optimal ``location'' of PMUs. Towards this end, we introduce the notion of average resistance distance and use it in conjunction with the graph of the connectivity matrix $\bm{B}$. Lastly, in \secref{subsec:information_theory}, we comment on the analysis presented in \cite{Li2013}, which addressed the PMU placement problem in an information-theoretic setting.

\subsection{Average resistance distance}\label{subsec:average_resdistance}
The connectivity matrix $\bm{B}$, defined by \eqref{eq:adjcency_matrix}, reveals interesting insights into the minimum number of PMUs and their location in the power grid. The entries of $\bm{B}$ can be used to define a measure of average resistance distance of each bus to other buses in the system:
\begin{eqnarray}\label{eq:average_distance}
\lambda_i = \sum_{j=1}^{N}\frac{b_{ij}}{N-1}.
\end{eqnarray}
We define a vector $\bm{\lambda}=\left[\lambda_1,\dots,\lambda_N\right]$. Let $\lambda_{\min} = \min(\bm{\lambda})$. We claim that, if $\lambda_i > \lambda_{\min}$, a PMU need not be placed at the location of the $i^{\text{th}}$ bus. This claim is justified by the fact that, since $\lambda_i$ quantifies the amount of average electrical connectivity between the $i^{\text{th}}$ and other buses in the network, the higher the value of $\lambda_i$ lower is the necessity to place a PMU on that bus. We now reinforce this argument by conducting simulations.

\begin{figure}[h]
\centering
  \includegraphics[height=3in,width=3.5in]{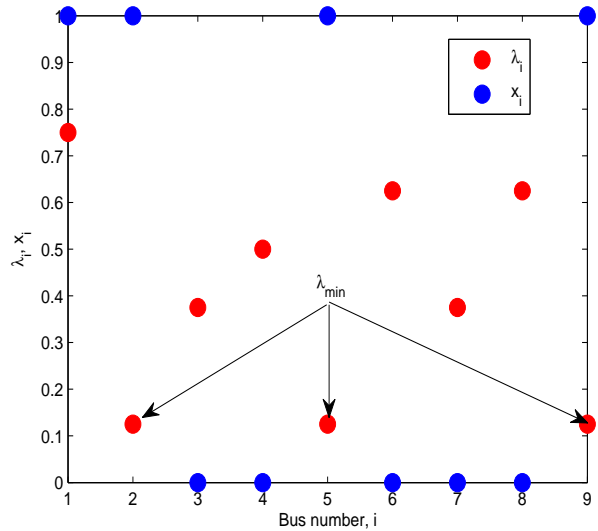}
  \caption{Average resistance distance for each bus for IEEE $9-$bus system.}
  \label{fig:lambda_x_9bus}
\end{figure}

From the connectivity matrix $\bm{B}$ defined by \eqref{eq:adjcency_matrix} for the IEEE $9-$bus system, and following \eqref{eq:average_distance}, we can plot $\lambda_i$ for each bus $i=1,\dots,9$ as shown in \figref{fig:lambda_x_9bus}. We also plot the optimal binary decision variable vector $\bm{x}$, obtained by solving \eqref{eq:pmuplacement} for the $9-$ system. We notice that, $x_i = 1$ ({\ie}, PMU to be installed) only when $\lambda_i = \lambda_{\min}$; for all other values of $\lambda_i$, $x_i = 0$ ({\ie}, no PMU). Furthermore, we can also infer the locations of the PMUs; they are to be installed on buses numbered $1$, $2$, $5$ and $9$. However, there exists a discrepancy: $x_1 = 1$ though $\lambda_1 > \lambda_{\min}$. This can be resolved by plotting the graph corresponding to the adjacency matrix $\bm{B}$, which we next describe.

\subsection{Graphical structure of the adjacency matrix $\boldsymbol{B}$}\label{subsec:graph_adjacency}
\begin{figure}
\centering
  \includegraphics[height=3in,width=3.5in]{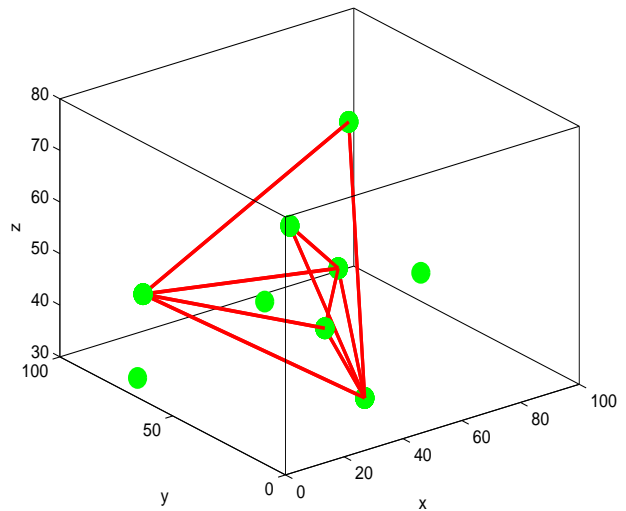}
  \caption{Graph corresponding to adjacency matrix $\bm{B}$ for IEEE $9-$bus system.}
  \label{fig:elec_9bus}
\end{figure}

The adjacency matrix can be visualized in terms of a graph, which can be used to further understand the PMU placement problem. We plot the graph corresponding to the adjacency matrix $\bm{B}$ for the IEEE $9-$bus system in \figref{fig:elec_9bus}. The points which are not connected in the graph correspond to buses with very low electrical connectivity to remaining buses in the network, which suggests that a PMU must be installed at these buses, while a single PMU is sufficient for the fully-connected subgraph of $\bm{B}$ to ensure complete network observability. Therefore, from \figref{fig:elec_9bus}, we infer that the total number of PMUs required for the $9-$bus system is $3+ 1 =4$, where $3$ corresponds to disconnected points, while $1$ corresponds to the fully-connected subgraph.
This discussion answers the ambiguity encountered in the previous subsection, {\ie}, if the fully-connected subgraph shown in \figref{fig:elec_9bus} corresponds to the point at which $\lambda_i > \lambda_{\min}$ in \figref{fig:lambda_x_9bus}, then a PMU can be installed at any one of the points of this fully-connected subgraph. Without loss of generality, we can install a PMU at bus numbered $1$ where we have $\lambda_1 > \lambda_{\min}$.

\begin{figure}[t]
\centering
  \includegraphics[height=3in,width=3.5in]{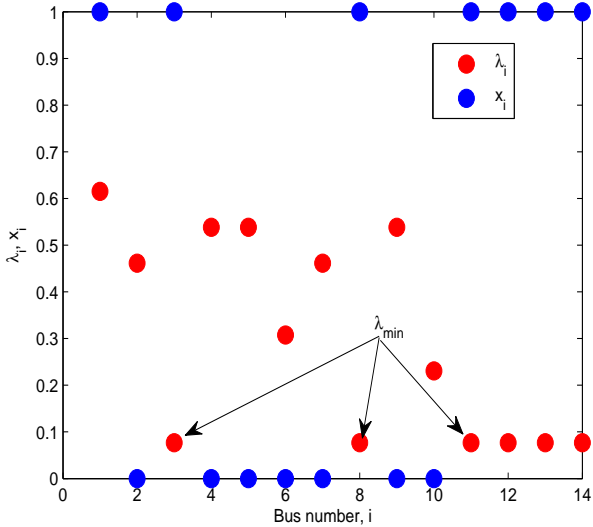}
  \caption{Average resistance distance for each bus for IEEE $14-$bus system.}
  \label{fig:lambda_x_14bus}
\end{figure}

\begin{figure}[t]
\centering
  \includegraphics[height=3in,width=3.5in]{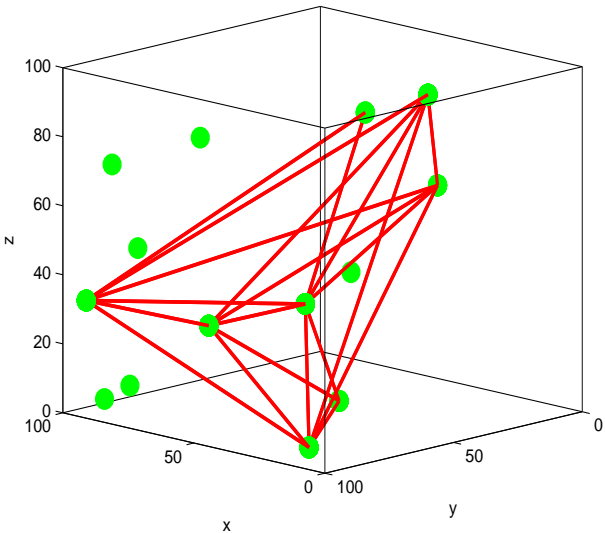}
  \caption{Graph corresponding to adjacency matrix $\bm{B}$ for IEEE $14-$bus system.}
  \label{fig:elec_14bus}
\end{figure}

Similar behavior was also noticed for the IEEE $14-$bus system, for which we present the average resistance distance and graph of the connectivity matrix in \figref{fig:lambda_x_14bus} and \figref{fig:elec_14bus}, respectively.
We see that the minimum required number of PMUs is $7$, and these PMUs must be placed at locations corresponding to the buses numbered $1$, $3$, $8$, $11$, $12$, $13$ and $14$. Buses numbered $i=3,8,11,12,13,14$ have $\lambda_i = \lambda_{\min}$, and correspond to disconnected points in the graph of matrix $\bm{B}$. Whereas, the PMU placed at bus numbered $1$ has $\lambda_1 > \lambda_{\min}$ and corresponds to the fully-connected subgraph of $\bm{B}$.

The minimum number of PMUs obtained using the average resistance distance and the graph of the adjacency matrix is in agreement with that obtained by solving \eqref{eq:pmuplacement}, with $\bm{A}=\bm{B}$ (see \tabref{tab:bus_pmu}). In the \appref{sec:appendix}, we show the plots for IEEE $30-$bus and $57-$bus systems. The plots for IEEE $118-$bus and $162-$bus systems are very dense, and are not shown here.

Interestingly, as claimed earlier, so far in this section we have addressed the PMU placement problem dealing only with the entries and the graph of matrix $\bm{B}$, without invoking the integer linear program \eqref{eq:pmuplacement}. This opens up the possibility of solving the placement problem without resorting to computationally intensive mathematical programming. These developments lead us to the following simple steps to obtain the minimum number of PMUs and address their location problem:
\begin{enumerate}
\item Derive the binary connectivity matrix, $\bm{B}$, using the electrical structure of the given bus network.
\item From the graph of $\bm{B}$, identify the fully-connected subgraph and the remaining disconnected points.
\item The minimum required number of PMUs = the number of disconnected points + $1$ (for the fully-connected subgraph).
\item Given $\bm{B}$, compute the average resistance distance $(\lambda_i)$ and the minimum average resistance distance $(\lambda_{\min})$ using \eqref{eq:average_distance}, for $i = 1,\dots,N$.
\item Install a PMU by default at bus $1$. The remaining PMUs are installed at locations where $\lambda_i = \lambda_{\min}$.
\end{enumerate}
The preceding discussion also suggests an intriguing connection between the average resistance distance and the graphical structure of the connectivity matrix, which is an open problem relegated to future work.

\begin{figure}[h]
\centering
  \includegraphics[height=3in,width=3.5in]{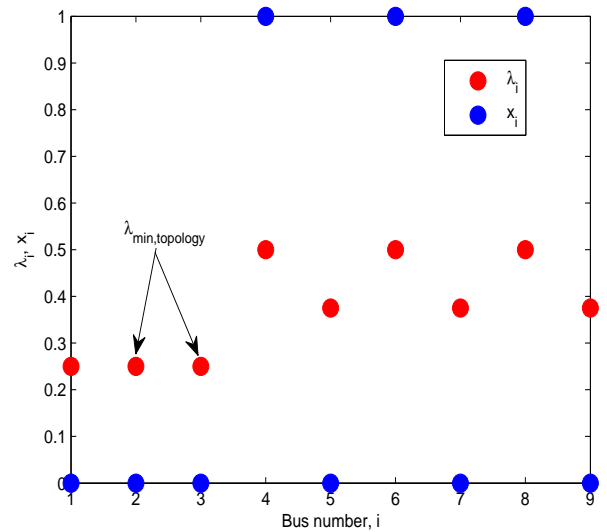}
  \caption{Average topological distance for each bus for IEEE $9-$bus system.}
  \label{fig:lambda_x_9bus_topo}
\end{figure}

\begin{figure}[h]
\centering
  \includegraphics[height=3in,width=3.5in]{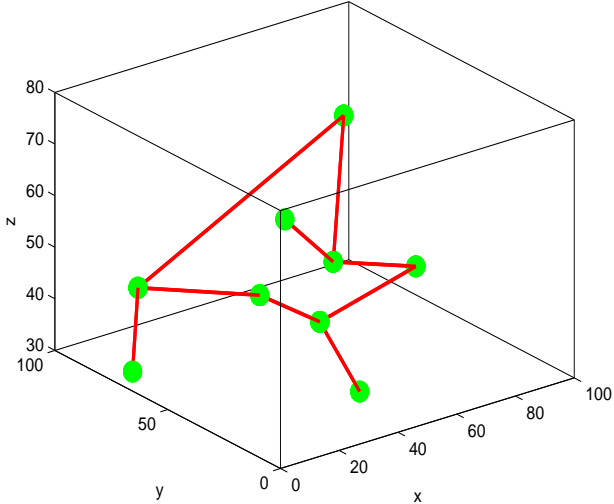}
  \caption{Graph corresponding to adjacency matrix $\bm{A}$ for IEEE $9-$bus system.}
  \label{fig:topo_9bus}
\end{figure}

Next, we plot the average topological distance and the graph of the connectivity matrix for the topological structure of IEEE $9-$bus system, in \figref{fig:lambda_x_9bus_topo} and \figref{fig:topo_9bus}, respectively. The average topological distance is given by
\begin{eqnarray}\label{eq:average_distance_topo}
\lambda_i = \sum_{j=1}^{N}\frac{a_{ij}}{N-1},
\end{eqnarray}
where $a_{ij}$'s are specified by \eqref{eq:topological_A}.
The discussions presented for the electrical structure do not hold for the topological structure of the power network. The plot involving the average topological distance does not reveal information pertinent to the location of the PMUs. Also, the graph representing the adjacency matrix $\bm{A}$ is not informative to evaluate the number of PMUs required for the bus network. Therefore, in the absence of zero injection measurements, the PMU placement problem based on the topological structure turns out to be a simple dominating set problem.

\subsection{Comments on \cite{Li2013}}\label{subsec:information_theory}
Highlighting the drawbacks of the topological observability criterion employed in conventional approaches, \cite{Li2013} proposed the use of an information-theoretic measure, namely, mutual information (MI) \cite{Cover2006} to address the PMU placement problem. The goal was to maximize the MI between PMU measurements and power system states to obtain highly ``informative'' PMU configurations. In the context of power systems, the ``information gain'' referred to the reduction in uncertainty of the power system states given the PMU measurements; the measurement set can also include zero power injections.

More precisely, let $\bm\theta$, $\boldsymbol{x}_{\text{PMU}}$ and $\boldsymbol{x}_{\text{conv}}$ denote vectors of voltage phasor angles at the buses, PMU measurements and conventional measurements, respectively. For a given PMU configuration $\mathcal{S}$ and $\boldsymbol{x}_{\text{conv}}$, the MI between $\bm{\theta}$ and $\boldsymbol{x}_{\text{PMU}}$ is given by
\begin{eqnarray}\label{eq:inftheorypmu1}
\nonumber I(\bm{\theta};\boldsymbol{x}_{\text{PMU}}(\mathcal{S})|\boldsymbol{x}_{\text{conv}}) = H(\bm{\theta}|\boldsymbol{x}_{\text{conv}})\\ - H(\bm{\theta}|\boldsymbol{x}_{\text{PMU}}(\mathcal{S}),\boldsymbol{x}_{\text{conv}}),
\end{eqnarray}
where $H(\bm{\theta}|\boldsymbol{x}_{\text{conv}})$ is the measure of uncertainty (entropy) of $\bm{\theta}$ given $\boldsymbol{x}_{\text{conv}}$. $H(\bm{\theta}|\boldsymbol{x}_{\text{PMU}}(\mathcal{S}),\boldsymbol{x}_{\text{conv}})$ is defined similarly. For sake of brevity, we neglect conventional measurements $\boldsymbol{x}_{\text{conv}}$. In \cite{Li2013}, the system states $\bm{\theta}$ were modeled as the Gaussian Markov random field (GMRF), for which the conditional entropy $H(\bm{\theta}|\boldsymbol{x}_{\text{PMU}}(\mathcal{S}))$ is given by
\begin{eqnarray}\label{eq:inftheorypmu2}
\nonumber H(\bm{\theta}|\boldsymbol{x}_{\text{PMU}}(\mathcal{S})) \approx \log\det \text{Cov}\left(\bm{\theta} - \hat{\bm{\theta}}\right) + \\ \frac{N}{2}\log (2\pi e) - N\log \delta,
\end{eqnarray}
where $\det$ and $\text{Cov}$ denote determinant and covariance, respectively; $N$ is the number of buses and $\delta$ is the quantization parameter. $\hat{\bm{\theta}}$ is the minimum mean square error (MMSE) estimate of $\bm{\theta}$, given $\boldsymbol{x}_{\text{PMU}}(\mathcal{S})$.

In \cite{Li2013}, it is claimed that, since $H(\bm{\theta})$ is fixed, the maximization of MI given by \eqref{eq:inftheorypmu1} is equivalent to minimization of the state estimation error given by $\log\det \text{Cov}\left(\bm{\theta} - \hat{\bm{\theta}}\right)$. This is central to the analysis and results presented there. However, this equivalence was not mathematically proved. In fact, it was shown in \cite{Guo2005} that MI and MMSE satisfy the following relationship for any distribution:
\begin{eqnarray}\label{eq:mi_mmse}
\frac{d}{d \mathrm{SNR}}\mathrm{I}(\mathrm{SNR}) = \frac{1}{2}\mathrm{MMSE}(\mathrm{SNR}),
\end{eqnarray}
where $\mathrm{SNR}$ is the signal-to-noise ratio. Therefore, given that \eqref{eq:mi_mmse} was rigorously proved in \cite{Guo2005}, the work presented in \cite{Li2013} needs to be revisited.

\section{Conclusion}\label{sec:conclusion}
In this paper, we revisited the PMU placement problem, but from the perspective of the electrical structure of the power network. The binary connectivity matrix used in the integer linear program formulation was derived using the resistance distance, which takes into account the underlying physical laws, namely, Ohm's and Kirchoff's. The problem formulated in this manner was tested on several standard IEEE test bus systems for complete network observability without zero injection measurements. The main result of the paper was that more number of PMUs were needed compared with those obtained using the topological structure of the network. However, given that the electrical structure provides a more comprehensive description of interconnections between components in the power network, our results promotes a realistic perspective of PMU placement in electric power systems. We explored the connectivity matrix, obtained using the topological structure, to provide insights into location of PMUs, and suggested avenues to address the placement problem without resorting to the computationally intensive optimization setup.

Future work would involve, among other things,
\begin{enumerate}
\item considering incomplete network observability, where the full set of system states (bus voltages) can be determined using lesser number of PMUs than obtained for complete observability. This would involve a study on the influence of the electrical structure on the depth-of-unobservability (first introduced in \cite{Nuqui2005});
\item including zero injection measurements, and verify if these measurements affect the optimal placement; and
\item to derive the fundamental principles to establish the connections between the average resistance distance and the graph of the adjacency matrix obtained using the electrical structure of the power grid.
\end{enumerate}

\appendices
\section{}\label{sec:appendix}
Here, we show the plots of the average resistance distance and the graph of the adjacency matrix $\bm{B}$ for IEEE $30-$bus and $57-$bus systems.
\begin{figure}[h]
\centering
  \includegraphics[height=3in,width=3.5in]{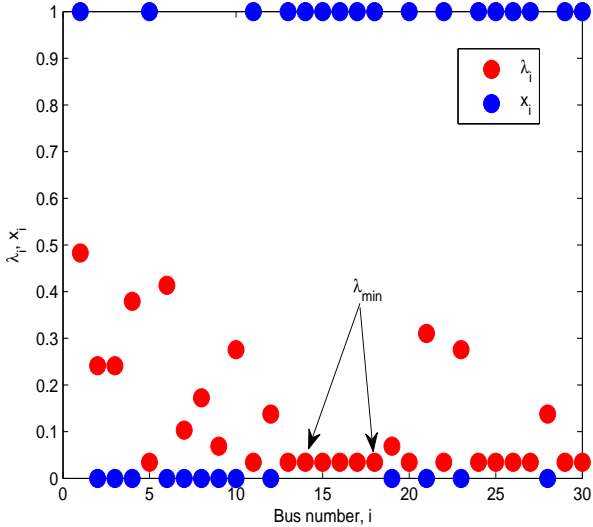}
  \caption{Average resistance distance for each bus for IEEE $30-$bus system.}
  \label{fig:lambda_x_30bus}
\end{figure}
\begin{figure}[h]
\centering
  \includegraphics[height=3in,width=3.5in]{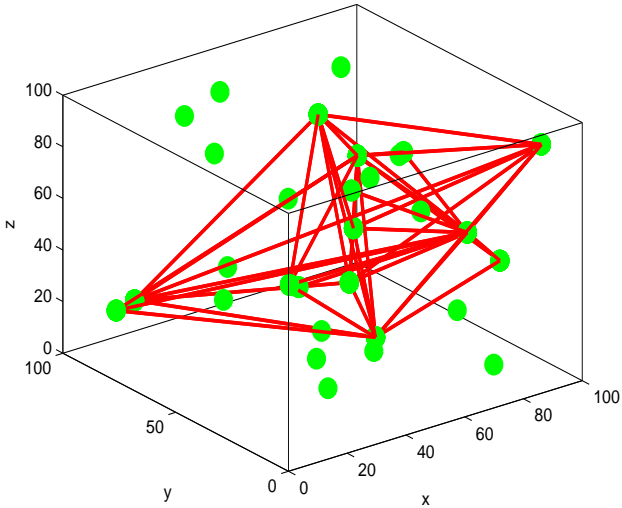}
  \caption{Graph corresponding to adjacency matrix $\bm{B}$ for IEEE $30-$bus system.}
  \label{fig:elec_30bus}
\end{figure}
For the $30-$bus system, we see from \figref{fig:lambda_x_30bus} that there are $16$ bus locations with $\lambda_i = \lambda_{\min}$ where a PMU has to be installed along with a PMU at bus $1$, resulting in a minimum number of $17$ PMUs for complete network observability. This is in full agreement with the number obtained by solving \eqref{eq:pmuplacement}, with $\bm{A}=\bm{B}$ (see \tabref{tab:bus_pmu}). Furthermore, this also corroborates with the $16$ disconnected points in \figref{fig:elec_30bus} plus one fully-connected subgraph.

\begin{figure}[h]
\centering
  \includegraphics[height=3in,width=3.5in]{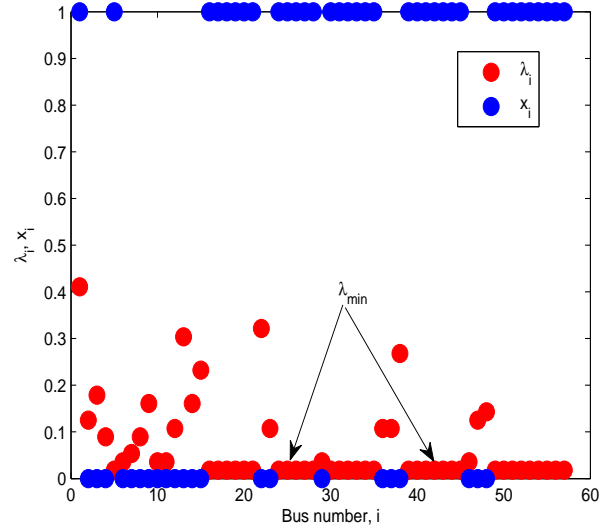}
  \caption{Average resistance distance for each bus for IEEE $57-$bus system.}
  \label{fig:lambda_x_57bus}
\end{figure}

Similar inferences can be drawn for the $57-$bus system whose plots are shown in \figref{fig:lambda_x_57bus} and \figref{fig:elec_30bus}.
\begin{figure}[h]
\centering
  \includegraphics[height=3in,width=3.5in]{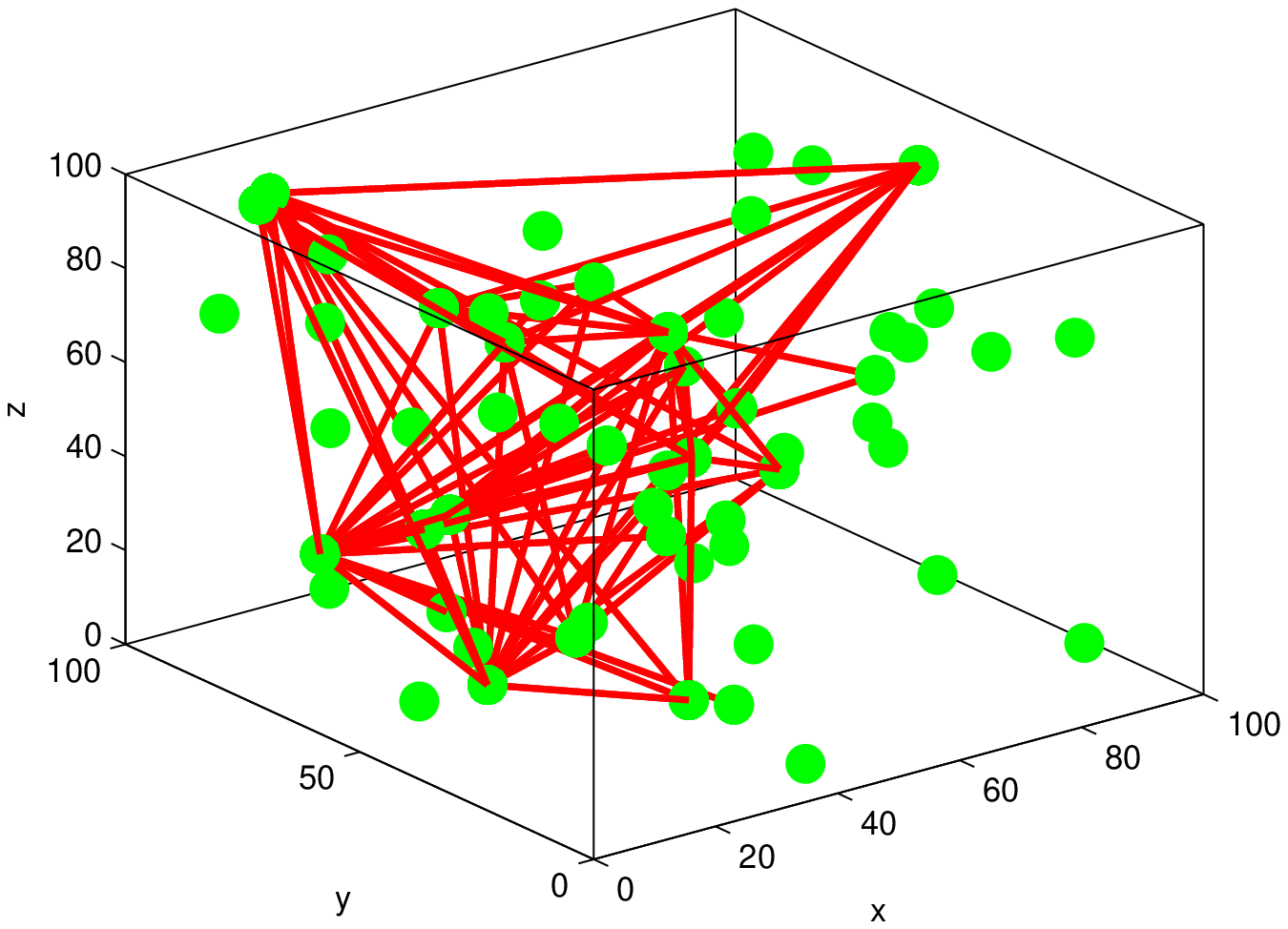}
  \caption{Graph corresponding to adjacency matrix $\bm{B}$ for IEEE $57-$bus system.}
  \label{fig:elec_57bus}
\end{figure}

\bibliographystyle{IEEEtran}
\bibliography{IEEEabrv,powersystems}

\end{document}